\documentclass[12pt,pdflatex]{article}
\usepackage{graphicx}
\evensidemargin \oddsidemargin
\usepackage{amsmath,amssymb,amsfonts,color,graphicx,cite,color,soul}
\usepackage{epsfig}
\usepackage[utf8]{inputenc}
\pdfoutput=1
\usepackage[T1]{fontenc}
\usepackage{latexsym}
\usepackage{mathrsfs}
\usepackage{dcolumn}
\usepackage{bm}
\usepackage[bottom]{footmisc}
\usepackage[affil-it]{authblk}
\usepackage[usenames]{xcolor}
\usepackage[english]{babel}
\usepackage{cite}
\usepackage{dcolumn}
\usepackage{color}

\newcommand{\sh}[1]{#1\hskip-7pt \diagup}

\begin{document}

\title{Lepton flavor violation in an extended MSSM}

\author[1]{R. Espinosa-Casta\~neda\footnote{rafael.espinosa.castaneda@cern.ch}}
\author[2]{F.V. Flores-Baez\footnote{fflores@fcfm.uanl.mx}}
\author[2]{M. G\'omez-Bock\footnote{melina.gomez@udlap.mx}}
\author[3]{M. Mondrag\'on\footnote{myriam@fisica.unam.mx}}

\affil[1]{\small CERN}
\affil[2]{\small Facultad de Ciencias Físico-Matemáticas, Universidad Autónoma de Nuevo León
Ciudad Universitaria, San Nicolás de los Garza, Nuevo León, 64450, México.}
\affil[3]{\small DAFM, Universidad de las Am\'ericas Puebla, Ex. Hda. Sta. Catarina M\'artir s/n, San Andr\'es Cholula, Puebla. México}
\affil[4]{\small Instituto de F\'isica, Universidad Nacional Aut\'onoma de M\'exico, Apdo. Postal 20-364, Cd. M\'exico 01000, M\'exico.}

\maketitle
\begin{abstract}
In this work we explore a lepton flavor violation  effect induced at one loop for a flavor structure  in an extended minimal standard supersymmetric model,
considering an ansatz for the trilinear term. In particular we find a finite expression which will show the impact of this phenomena in the  $h\to \mu \tau$ decay,  
produced by  a mixing in the trilinear coupling of the soft supersymmetric Lagrangian. 
\end{abstract}

\section{Introduction}

In the Standard Model (SM) lepton flavor violation processes were forbidden by the lepton number conservation,
which is not associated with a gauge symmetry. In
the SM, the spontaneous breaking of the electroweak
symmetry produces eigenstates of the remaining gauge group that are
not in general eigenstates of the mass matrix
\cite{Weinberg:1967tq,Weinberg:1972tu,Glashow:1961tr,Glashow:1970gm}. But
after diagonalization of the mass matrix, the electroweak coupling matrix is also
diagonal in the mass basis, therefore there is no possibility for
lepton flavor violation. Nevertheless, this symmetry is lost in neutrino oscillation found in experiments 
\cite{Cleveland:1998nv,Fukuda:1998mi,Ahmad:2002jz,Ahn:2002up}, which forces the model structure to go beyond the SM. 

This lost symmetry on leptons observed in the neutrino mixing evidence opens also the possibility of lepton flavor violation in the charged sector.
Experimental data taken from CMS at $8~TeV$ with $19.7fb^{-1}$ had shown an excess for $BR(h^0\to \tau \mu)$ of $2.4\sigma$ 
with best fit branching fraction of $0.84^{+0.39}_{-0.37}\%$ \cite{Khachatryan:2015kon}.
Nevertheless data from ATLAS has shown only a $1\sigma$ significance for the same process \cite{Aad:2016blu}. 
Moreover, also recent measurements at $13~TeV$, although with only $2.3fb^{-1}$ of data, has shown no evidence of excess.
It was even reported \cite{CMS-PAS-SUS-15-006} a best fit branching fraction of $-0.56\%$. The sign change  may imply a statistical error of the data.
Even for those latest reports, if they are confirmed, they will indicate a very low range for this lepton flavor violation processes to occur at these energies. 
These  will set stringent bounds to any model beyond the SM. One of the works toward this direction is done in \cite{Sierra:2014nqa}, 
where it has been explored it in a Two Higgs Doublet Model
with flavor violation in the Yukawa couplings, a model known as THDM-III \cite{GomezBock:2005hc}.
 In order to fit these restrictions bounds, the model we proposed in this work of a scalar flavor extended MSSM, 
should show exclusions regions in the parameter space.  \\

The most recent data by the LHC at $13~TeV$ have not shown evidence for supersymmetry  for different channels and observables as events with one lepton as final state
\cite{CMS-PAS-SUS-15-006}, jets and leptons or three leptons \cite{Aad:2016tuk}, missing energy \cite{Bradmiller-Feld:2016yvd}. 
The experimental search is mainly for the Lightest Supersymmetric Particle (LSP) as missing energy, these reports analyses the data for simplified supersymmetric 
models at this specific energy. The results of these analysis have reduced the parameter space for Minimal Supersymmetric Standard Model (MSSM). 
Nevertheless we claim that is important to fully explore the possible parameter space for different non-simple supersymmetric low energy models knowing that
Supersymmetry still solves many phenomenological issues \cite{Haber:2000jh,Heinemeyer:2011aa} and also is needed for many GUT models. 
A review on flavor violation processes in the MSSM considering neutrino mixing can be found in \cite{Vicente:2015cka}. 
In this work we implement a flavour structure which implies flavour violation and non-universality of sfermions, 
this was previously introduced in \cite{GomezBock:2008hz} and was used also in a similar work in \cite{Aloni:2015wvn}.

\section{Flavor mixed sleptons}

As the evidence on flavor violation in charged lepton is not yet conclusive but gives low values for these branching ratios, 
one possibility is to have a mixed flavor structure 
in an unseen sector as the sfermions, and then this mixing will induce flavor violation through radiative corrections.\\
The mixing of slepton states will be given by a flavor structure in the mass matrix, specifically we explore the possibility of having mixing flavor terms in 
the trilinear couplings of the soft supersymmetric Lagrangian. 
\begin{equation}
\mathcal{L}_{soft}^{f}=-\sum_{\tilde f_i} \tilde{M}_{\tilde{f}}^{2}\tilde{\bar{f}}_{i}\tilde{f}_{i}
-(A_{\tilde{f},i}\tilde{\bar{f_L}}^{i}H_{1}\tilde{f_R}^{i}+h.c),
\label{LsoftNFV}
\end{equation}
where $\tilde{f}$ are the scalar fields in the supermultiplet. In the
case of sfermions, as they are scalar particles, the $L,R$ are just labels which point out to the
fermionic SM partners, but they no longer  have left and right $SU(2)$ properties. In general they may
mix in two physical states.\\
Once the EW symmetry breaking is considered, the trilinear term of the soft SUSY Lagrangian
 for the sleptonic sector takes the following form
\begin{equation}
  \mathcal{L}_{H\tilde{f}_{i}\tilde{f}_{j}}=
  \frac{A_{l}^{ij}}{\sqrt{2}}\left[(\phi_{1}^{0}-i\chi_{1}^{0})\tilde{l}_{iR}^{*}\tilde{l}_{jL}-\sqrt{2}\phi_{1}^{-}\widetilde{l}_{iR}^{*}\widetilde{\nu}_{jL}
    +v_{1}\tilde{l}_{R}^{*}\tilde{l}_{L}\right] + h.c.  \nonumber 
\end{equation}
The contribution to the elements of the sfermion mass matrix come from
the interaction of the Higgs scalars with the sfermions, which appear
in different terms of the superpotential and soft-SUSY breaking terms
as is fully explained in \cite{Kuroda:1999ks,Okumura:2003hy}. In the
case of the slepton mass matrix, as we said before, the contributions
coming from {\it mass soft terms} are $\tilde{M}_{l,LL}^{2}$,
$\tilde{M}_{l,RR}^{2}$, from trilinear couplings after EW symmetry
breaking $A_{ij}^l$ and from the $F,D -$terms. We arrange them in
a block mass matrix as follows
\begin{equation}
\tilde{M}_{l}^{2}=
\begin{pmatrix}
    m_{LL,l}^{2} &  m_{LR,l}^{2} \\
    m_{RL,l}^{2} & m_{RR,l}^{2}
\end{pmatrix}~.
\label{massdFD}
\end{equation}
The elements of the sleptons mass matrix eq. (\ref{massdFD}), for the different flavors 
given by $i,j=e,\mu,\tau$ 
are 
\begin{eqnarray}
 m_{LL,l}^{2}& =&
\tilde{M}_{\tilde{L,l}}^{2}+m_{l_{L}}^{2}+\frac{1}{2}\cos2\beta(2M_{W}^{2}-M_{Z}^{2}),\\
m_{RR,l}^{2} & = &
\tilde{M}_{\tilde{E},l}^{2}+m_{l_{R}}^{2}-\cos2\beta\sin^{2}\theta_{W}M_{Z}^{2},\\
m_{LR,l}^{2} & = &
\frac{A_{l}v\cos\beta}{\sqrt{2}}-m_{l}\mu_{susy}\tan\beta.
\label{Aterm}
\end{eqnarray}
The soft terms are not the only contributions to the sfermion mass elements, the supersymmetric auxiliary fields $F$ and $D$ coming from the superpotential
also contribute to this mass matrix.\\
In the present work, we assume the mass terms coming from SUSY breaking terms dominate over the EW terms coming from auxiliary $F$ and $D$ fields so we can safely 
approximate the diagonal elements to a soft susy scale $m_S$ as
$m_{RR}^{2} \approx
m_{LL}^{2} =m_{S}^{2} {\bf 1_{2\times 2}}$.\\
In order to analyze the consequences of a flavor structure we construct an ansatz for the trilinear terms $A_t$. Our procedure is similar to
the work done in Ref. \cite{DiazCruz:2001gf} for FCNC's in the quark sector through an ansatz of soft-susy terms. 
In our case we consider the whole two families contributions 
and of the same order of magnitude, having the following form for the trilinear term
\begin{equation}
 A_{l}=
\begin{pmatrix}
  0 & 0 & 0 \\
  0 & w & y \\
  0 & y & 1
\end{pmatrix}
A_{0} \label{BLO} .
\end{equation}
In this case, one could have at tree level the selectrons in a singlet
irreducible representation decoupled from the other two families of sleptons.  This would
give rise to a $4\times 4$ matrix, diagonalizable through a unitary
matrix $Z_{\tilde{l}}$, such that
$Z_{\tilde l}^{\dag}\tilde{M}_{l}^{2}Z_{\tilde l}=\tilde{M}_{diag}^{2}$. \\
We will have then  physical non-degenerate slepton masses\footnote{We assign the label $\tilde{\tau},\tilde{\mu}$ 
to the masses to show the relation to the non-FV sleptons.}
\begin{align}
m^{2}_{\tilde{\tau}_{1,2}}& = \frac{1}{2}(2 \tilde{m}_S^2-X_{m}-X_{t}\pm R)\ ,\nonumber\\
m^{2}_{\tilde{\mu}_{1,2}}& =  \frac{1}{2}(2 \tilde{m}_S^2+X_{m}+X_{t}\pm R)\ ,
\end{align}
where $R=\sqrt{4 A_y^2+\left(X_{t}-X_m \right)^2}$ with  $A_{y}=\frac{1}{\sqrt{2}}yA_{0}v\cos\beta$, 
$X_{m}=\frac{1}{\sqrt{2}}wA_{0}v\cos\beta - \mu_{susy} m_{\mu}\tan\beta$
and $X_{t}=\frac{1}{\sqrt{2}}A_{0}v\cos\beta - \mu m_{\tau}\tan\beta$. \\
We may  write the transformation which diagonalizes the mass matrix as a $4\times4$ rotation matrix for sleptons 
$Z_{\tilde{l}}$, which is in turn  a $2\times2 $ block matrix
$ Z_{\tilde l}^{\dag}\tilde{M}_{\mu -\tau}^{2} Z_{\tilde l}=\tilde{M}_{l,diag}^{2}$, explicitly having
 the form
\begin{equation}\label{rotationZ}
Z_{\tilde{l}}=\frac{1}{\sqrt{2}}\left(
\begin{array}{cc}
\Phi & -\Phi\\
\Phi\sigma^3 & \Phi\sigma^3
\end{array}
\right) ~,
\end{equation}
\vspace{1cm}
where $\sigma_3$ is the Pauli matrix and
\begin{eqnarray}
\Phi&=\left(
\begin{array}{cc}
-\sin\frac{\varphi}{2} & -\cos\frac{\varphi}{2}\\
\cos\frac{\varphi}{2} & -\sin \frac{\varphi}{2}
\end{array}
\right),\,\,\,\,
\tan\varphi  = \frac{2A_y}{X_m -X_t},\nonumber \\
\end{eqnarray}

The non-physical states are transformed to the physical eigenstates as
\begin{equation}
\begin{pmatrix}
\tilde{\mu}_{L} \\
  \tilde{\tau}_{L}\\
  \tilde{\mu}_{R}\\
  \tilde{\tau}_{R}
\end{pmatrix}
= Z_{\tilde l}
\begin{pmatrix}
  \tilde{l}_{1}\\
  \tilde{l}_{2}\\
  \tilde{l}_{3}\\
  \tilde{l}_{4}
\end{pmatrix}
 \label{rotationB}
\end{equation}

\subsection{Higgs flavor violation coupling with sleptons}

The Lagrangian which gives the interaction of scalar neutral light Higgs $h^0$-slepton-slepton is given as
\begin{eqnarray}
 \mathcal{L}_{h^0\tilde{l}\tilde{l}}= Q_{\tilde{l}}\left[\tilde{l}_L^*\tilde{l}_L+\tilde{l}_R^*\tilde{l}_R\right]h^0
 +G\left[(-\frac{1}{2}+s_w^2)\tilde{l}_L^*\tilde{l}_L-s^2_w \tilde{l}_R^*\tilde{l}_R\right]h^0+\chi_{\tilde{l}}\left[\tilde{l}_L^*\tilde{l}_R+\tilde{l}_R^*\tilde{l}_L \right]
\end{eqnarray}
where 
\begin{eqnarray}
 &Q_{\tilde{\mu},\tilde{\tau}}&=\frac{gm_{\tilde{\mu},\tilde{\tau}}^2\sin\alpha}{M_w\cos\beta}\ ,\\
 &G&=g_zM_z\sin(\alpha + \beta)\ ,\\
 &X_{\tilde{\mu},\tilde{\tau}}&=\frac{gm_{\tilde{\mu},\tilde{\tau}}\sin\alpha}{2M_wcos\beta}(A_{\tilde{\mu},\tilde{\tau}}-\mu \cot\alpha)\ .
\end{eqnarray}
Then, in the slepton physical states, as rotated by (\ref{rotationB}), the couplings to the light Higgs boson are given as in table \ref{table:V3}, 
where $T_{\tilde{l}}^{\pm}=Q_{\tilde{l}}\pm X_{\tilde{l}}$.
We have simplified the notation using $s_{\varphi}=\sin\frac{\varphi}{2}$ and $c_{\varphi}=\cos\frac{\varphi}{2}$.\\
{\tiny
\begin{table}
\begin{center}
\resizebox{14cm}{!}{
\begin{tabular}{lcccc}
\hline
$g_{h^0\tilde{f}\tilde{f}}$&$\tilde{\mu}_1$&$\tilde{\mu}_2$&$\tilde{\tau}_1$&$\tilde{\tau}_2$\\
\hline
\hline
$\tilde{\mu}_1$& $s_{\varphi}^2 T_{\tilde{\tau}}^+ + c_{\varphi}^2T_{\tilde{\mu}}^+ -\frac{1}{4}G $&$\frac{1}{4}G(1-4s_w^2)$&0&
$s_\varphi c_\varphi(T_{\tilde{\tau}}^+-T_{\tilde{\mu}}^+)$\\
\hline
$\tilde{\mu}_2$&*& $s_{\varphi}^2T_{\tilde{\tau}}^- + c_{\varphi}^2T_{\tilde{\mu}}^- -\frac{1}{4}G$&
$s_{\varphi}c_{\varphi}(T_{\tilde{\tau}}^--T_{\tilde{\mu}}^-)$
&0\\
\hline
$\tilde{\tau}_1$&0&*&$s_{\varphi}^2T_{\tilde{\mu}}^- + c_{\varphi}^2T_{\tilde{\tau}}^--\frac{1}{4}G$&$\frac{1}{4}G(1-4s_w^2)$\\
\hline
$\tilde{\tau}_2$&*&0&*& $s_{\varphi}^2T_{\tilde{\mu}}^+ + c_{\varphi}^2T_{\tilde{\tau}}^+-\frac{1}{4}G$ \\
\hline
\hline
\end{tabular}
}
\end{center}
\caption{\label{table:V3}Expressions of the respective interactions of the Higgs boson $h^0$ with two sleptons}
\end{table}
}

\section{Radiative Higgs flavour violation decay} 
In this section we show the construction of the  radiative induced Higgs flavor violation decay. We assume the following convention:  
the slepton which interacts with the muon (tau) is labeled  with the index \textit{i} (\textit{j}), see Fig 1. 
 \begin{figure}[hbt!]
 \label{h0fvDiag}
 \begin{center}
\includegraphics[height=3cm]{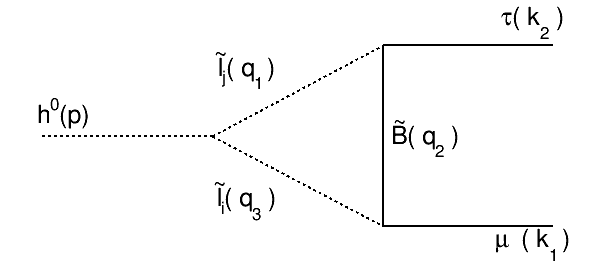}
\caption[]{\it 1-loop SUSY slepton flavor mixing contribution to $h^0 \to \mu \tau$.}
\end{center}
\end{figure}
The notation used for the coupling between  the bino $\tilde{B}$, the slepton $\tilde{l}$ and the lepton $l$, for  $l=\mu, \tau$,
which is denoted  as $\tilde{B}\tilde{l} l$, can be written in terms of three types of coefficients for each lepton. 
We use $a_k$, for the ones in the coupling with the muon 
and $b_k$ for coefficients in the coupling with the tau; with  $k=1,2,3$; then we have this couplings written as:
\begin{align}
g_{\tilde{B}\tilde{l}_i\mu}&=a^{i}_{1}[a^{i}_{2}+a^{i}_{3}\gamma^{5}] \ ,\nonumber \\
g_{\tilde{B}\tilde{l}_j\tau}&=b^{j}_{1}[b^{j}_{2}+b^{j}_{3}\gamma^{5}]\  ,
\label{gBsll}
\end{align}
where the first type of coefficients are given as 
\begin{align}
a^{i}_{1}&= \frac{g}{4}\tan\theta_{W}a_{\tilde{l}_i}\ ,\\
b^{j}_{1}&= -\frac{g}{4}\tan\theta_{W}b_{\tilde{l}_j}\ ,
 \end{align}
being $a_{\tilde{l}_i}= s_{\varphi}(-c_{\varphi})$ for $\tilde{\mu}_{1,2}( \tilde{\tau}_{1,2} )$, while
$b_{\tilde{l}_j}= c_{\varphi}, (s_{\varphi})$ for $ \tilde{\mu}_{1,2}( \tilde{\tau}_{1,2} )$.
And the rest of the coefficients, type 2 and 3 are  numbers shown in table \ref{ab23}. \\

\begin{table}
\caption{\label{ab23}Coefficients for $a_{2,3}$ and $b_{2,3}$ of Bino-slepton-lepton couplings $g_{\tilde{B}\tilde{l}l}$ given in eq. (\ref{gBsll})}
\begin{center}
\resizebox{8cm}{!}{
\begin{tabular}{cccccccccc}
\hline
\hline
 & $i,j=1$ & $2$ & $3$ & $4$& &$1$ &$2$&$3$&$4$ \\ 
\hline
\hline
$a_{2}^{i}$ & $1$ &$3$ &$3$ & $1$&$a_3^{i}$& $3$ &$1$&$1$&$3$ \\ 
\hline 
$b_{2}^{j}$ & $1$ &$3$ &$3$ & $1$&$b_3^{j}$& $3$ &$1$&$1$&$3$ \\ 
\hline
\hline
\end{tabular} }
\end{center}
\end{table}
Now we write the invariant amplitude for the $h^{0}\to \mu \tau$ decay with   slepton $\tilde{l}_{i}, \tilde{l}_{j} $ running inside the loop as indicated in lines above,  
in such a way that it is possible to identify every vertex, propagator and external fermion legs, see figure \ref{h0fvDiag}
\begin{align}\label{eq:general}
\mathcal{M}_{ij}&= \int \frac{d^4q}{(2\pi)^4} \bar{v}(k_{1}) [ \imath  g_{\tilde{B}\tilde{l}_{i}\mu} ]\left [\frac{\imath(\sh q-\sh k_{2}+m_{B})}{(q-k_{2})^2-m^2_{B}}\right][ \imath  g_{\tilde{B}\tilde{l}_{j}\tau} ]\left[\frac{\imath}{(q-p)^2-m^2_{i}}\right] \left[\frac{\imath}{q^2-m^2_{j}}\right] [\imath g_{h\tilde{l}_i\tilde{l}_{j}}]u(k_{2})
\end{align}
After the  loop momentum integration over $q=q_1$, the amplitude can be written in the following form
\begin{align}
\mathcal{M}_{ij}&= \bar{v}(k_{1})[A^{ij} + B^{ij} \gamma^{5}] u(k_{2})\ ,
\end{align}
where
\begin{align}\label{eq:aij}
A^{ij}&= a_{1}^{i}b_{1}^{j}\big[ (a_{2}^{i}b_{2}^{j}+a_{3}^{i}b_{3}^{j})m_{B}F^{ij}
 + (-a_{2}^{i}b_{2}^{j}+a_{3}^{i}b_{3}^{j})m_{\tau}F^{ij} \nonumber\\ &
 + (a_{2}^{i}b_{2}^{j}- a_{3}^{i}b_{3}^{j})m_{\tau}F^{ij}_{II} - (a_{2}^{i}b_{2}^{j}- a_{3}^{i}b_{3}^{j})m_{\mu}F^{ij}_{III}\big]\ ,\\
\label{eq:bij}B^{ij}&=a_{1}^{i}b_{1}^{j} \big[ (a_{2}^{i}b_{3}^{j}+a_{3}^{i}b_{2}^{j})m_{B}F^{ij} 
- (a_{3}^{i}b_{2}^{j} - a_{2}^{i}b_{3}^{j})m_{\tau}F^{ij}\nonumber\\  
&+ (a_{3}^{i}b_{2}^{j} - a_{2}^{i}b_{3}^{j})m_{\tau}F^{ij}_{II} - (a_{2}^{i}b_{3}^{j} - a_{3}^{i}b_{2}^{j})m_{\mu}F^{ij}_{III}\big]\ .
\end{align}
The $F$ functions are given as
\begin{align}
F^{ij}&=-g_{h\tilde{l}_{i}\tilde{l}_{j}}\frac{\imath}{16\pi^2}C0[m^2_{H},m^2_{\mu},m^2_{\tau},m^2_{j},m^2_{i},m^2_{B}]\ ,\\
F^{ij}_{II}&=-g_{h\tilde{l}_{i}\tilde{l}_{j}}\frac{\imath}{(16\pi^2)[m^2_{H} -(m_{\mu}+m_{\tau})^2 ][m^2_{H} -(m_{\mu}-m_{\tau})^2]} \nonumber\\
&\times \big[ -C0(m^2_{H},m^2_{\mu},m^2_{\tau},m^2_{j},m^2_{i},m^2_{B})\big[m^2_{B}(m^2_{H}+m^2_{\mu}-m^2_{\tau})\nonumber\\
&+m^4_{H} + m^2_{\mu}(m^2_{i}-2m^2_{j})-m^2_{H}(m^2_{i}+m^2_{\mu}+2m^2_{\tau})+m^2_{\tau}(m^2_{i}-m^2_{\mu})+m^4_{\tau}\big]\nonumber\\
&  -2m^2_{\mu}B0(m^2_{\mu},m^2_{B},m^2_{i}) + B0(m^2_{\tau},m^2_{B},m^2_{j})(m^2_{\mu}+m^2_{\tau}-m^2_{H})\nonumber\\
& -B0(m^2_{H},m^2_{i},m^2_{j})(m^2_{\tau}-m^2_{\mu}-m^2_{H})
\big]\ ,\\
F^{ij}_{III}&=-g_{h\tilde{l}_{i}\tilde{l}_{j}}\frac{\imath}{(16\pi^2)[m^2_{H} -(m_{\mu}+m_{\tau})^2 ][m^2_{H} -(m_{\mu}-m_{\tau})^2]} \nonumber\\
&\times \big[ -C0(m^2_{H},m^2_{\mu},m^2_{\tau},m^2_{j},m^2_{i},m^2_{B})\big[ m^2_{\tau}(m^2_{B}+m^2_{H}-2m^2_{i}+m^2_{j}+m^2_{\mu})\nonumber\\
& -m^4_{\tau}+(m^2_{B}-m^2_{j})(m^2_{H}-m^2_{\mu})\big]+2m^2_{\tau}B0(m^2_{\tau},m^2_{B},m^2_{j})\nonumber\\
&+B0(m^2_{\mu},m^2_{B},m^2_{i})(m^2_{H}-m^2_{\mu}-m^2_{\tau})
-B0(m^2_{H},m^2_{i},m^2_{j})(m^2_{H}-m^2_{\mu}+m^2_{\tau})
\big]\ .
\end{align}
Then the differential width for the process $h^0 \to \tau \mu$ can be calculated using the amplitude transition matrix as
\begin{align}
d\Gamma&= \frac{1}{32\pi^2}\sum_{spin}|\mathcal{M}_{T}|^2\frac{|p|}{m^2_{H}}d\Omega,\\
|p|&=\sqrt{(m_{H}^2-(m_{\tau}-m_{\mu})^2)(m_{H}^2-(m_{\tau}+m_{\mu})^2)}\ .
\end{align}

Notice that the total amplitude includes all possible combinations of sleptons in the internal lines, then
\begin{align}
\mathcal{M}_{T}&=\sum_{i,j}\mathcal{M}_{ij}. \nonumber\\
\end{align}

\section{Conclusions}
In this work we consider a flavor structure on trilinear soft terms, 
assuming a two family mixing in the sleptons we explore the consequences of this structure in a particular process involving lepton flavor violation for the Higgs boson.
We obtain non-degenerate slepton masses for four of the sleptons which are decoupled from the first family and mixed in flavor. We also found that in physical basis two specific couplings of the Higgs boson with sleptons are zero, {\it i.e.}
$g_{h^0\tilde{\mu}_1\tilde{\tau}_1}=g_{h^0\tilde{\mu}_2\tilde{\tau}_2}=0$.\\
We obtain the expression for the one-loop radiative correction of the specific process $h^0\to \tau \mu$.
The expression we obtain is found to be  UV-finite and can be used to restrict the parameter space of the supersymmetric model applied to this process, as 
it is very restricted by the experimental data. 
This kind of structure also gives extra contribution to $BR(\tau \to \mu \gamma)$ and to the muon anomalous magnetic moment $g-2$.
So a complete exploration of the parameter for all these processes will be a goal for a further work.
 
\section*{Acknowledgements}
This work was partially supported by a Consejo Nacional de Ciencia y Tecnolog\'ia (Conacyt), Posdoctoral Fellowship
and SNI M\'exico. R.~E-C wants to thank CONACyT RED-Física de Altas Enerías. F.~F-B was partially supported by Universidad Aut\'onoma de Nuevo Le\'on through the grant PAICYT-2015. 
M.~G-B ac\-know\-led\-ges partial support from Universidad de las Am\'ericas Puebla. This work was also partially supported by grants UNAM PAPIIT IN111115 and
Conacyt 132059.

\bibliographystyle{iopart-num}
\bibliography{proh0LFV}

\end{document}